\documentclass[sigconf]{acmart}
\usepackage{hyperref}
\usepackage{tabularx}
\usepackage{longtable}
\usepackage[utf8]{inputenc}
\usepackage{listings}
\usepackage{colortbl}%

\usepackage{booktabs}
\usepackage{graphicx}
\newtheorem{mydef}{Definition}

\usepackage{cleveref}
\crefname{mydef}{Definition}{Definitions}
\newcommand{\ep}{\varepsilon}
\newcommand{\eptext}{$\ep$}
\newcommand{\simenv}{{\textsc{Simulation Environment}}}
\newcommand{\polleditor}{{\textsc{Poll Editor}}}
\newcommand{\Randori}{{\textsc{Randori}}}
\newcommand{\dpcomp}{{\textsc{DPComp}}}
\newcommand{\dpbench}{{\textsc{DPBench}}}
\newcommand{\alts}{{\textsf{alts}}}
\newcommand{\truth}{{\textsf{truth}}}
\newcommand{\weight}{{\textsf{weight}}}
\newcommand{\answers}{{\textsf{answers}}}
\newcommand{\depth}{{\textsf{depth}}}
\newcommand{\pop}{{\textsf{pop}}}
\newcommand{\tradeoff}{accuracy/privacy trade-off}

\renewcommand\footnotetextcopyrightpermission[1]{} % removes footnote with conference information in first column

\title{Efficient Error Prediction for Differentially Private Algorithms}

\author{Boel Nelson}
\affiliation{%
  \institution{Department of Computer Science and Engineering,\\ Chalmers University of Technology}
  \city{Gothenburg}\country{Sweden}
}\email{boeln@chalmers.se}

\begin{document}
\begin{abstract}
Differential privacy is a strong mathematical notion of privacy. Still, a prominent challenge when using differential privacy in real data collection is understanding and counteracting the accuracy loss that differential privacy imposes. As such, the \tradeoff\ of differential privacy needs to be balanced on a case-by-case basis. Applications in the literature tend to focus solely on analytical accuracy bounds, not include data in error prediction, or use arbitrary settings to measure error empirically.

To fill the gap in the literature, we propose a novel application of \textit{factor experiments} to create data aware error predictions. Basically, factor experiments provide a systematic approach to conducting empirical experiments. To demonstrate our methodology in action, we conduct a case study where error is dependent on arbitrarily complex tree structures. We first construct a tool to simulate poll data. Next, we use our simulated data to construct a least squares model to predict error. Last, we show how to validate the model. Consequently, our contribution is a method for constructing error prediction models that are data aware.
\end{abstract}

%   <concept>
%       <concept_id>10002978.10003018.10003019</concept_id>
%       <concept_desc>Security and privacy~Data anonymization and sanitization</concept_desc>
%       <concept_significance>500</concept_significance>
%       </concept>
%   <concept>
%       <concept_id>10002950.10003648.10003671</concept_id>
%       <concept_desc>Mathematics of computing~Probabilistic algorithms</concept_desc>
%       <concept_significance>300</concept_significance>
%       </concept>
%   <concept>
%       <concept_id>10010147.10010341.10010342.10010343</concept_id>
%       <concept_desc>Computing methodologies~Modeling methodologies</concept_desc>
%       <concept_significance>100</concept_significance>
%       </concept>
%  </ccs2012>

% \ccsdesc[500]{Security and privacy~Data anonymization and sanitization}
% \ccsdesc[300]{Mathematics of computing~Probabilistic algorithms}
% \ccsdesc[100]{Computing methodologies~Modeling methodologies}
\maketitle
\pagestyle{plain}

\section{Introduction}\label{paperE:sec:introduction}
Adopting differential privacy in real systems is ultimately an issue of properly understanding the impact differential privacy will have on accuracy. In other words, if an analyst cannot predict the accuracy loss an algorithm will cause, they will be hesitant to use the algorithm. As such, understanding the accuracy loss induced by differential privacy is crucial to differential privacy being deployed in real systems.

In the literature, the accuracy of a differentially private algorithm is often evaluated analytically through Chernoff bounds, such as by \citet{kasiviswanathan_what_2011}. Here, the authors introduce a metric for error, namely misclassification error, which is applicable in their domain. However, the general Chernoff bound they provide requires that there exists a definition for error, i.e. a unit of measurement for the inaccuracy introduced by differential privacy. As such, if the relationship between input variables and error is unknown, Chernoff bounds will not be applicable. As noted by \citet{hay_principled_2016}, the more complex algorithm, the more difficult it is to analyze the algorithm theoretically. Consequently, some algorithms may be easier to investigate empirically instead of analytically.

In addition, previous research~\cite{hay_principled_2016,hay_exploring_2016} shows that the accuracy of a differentially private algorithm may be greatly influenced by the input data. Consequently, input data should also be taken into account when modeling error. So far, the current literature seems to model error from the algorithm without taking the input data into consideration. For example, \citet{kasiviswanathan_what_2011} and \citet{vadhan_complexity_2017} use Chernoff bounds, but they do not include input data in their error model.

From the other end of the perspective, several papers including~\cite{ding_differentially_2011-4,xiao_dpcube_2012-1, li_differentially_2015-9,chen_private_2015, benkhelif_co-clustering_2017-3,gao_dynamic_2018, li_ihp_2019} investigate error empirically. Still, input values to the experiments are chosen seemingly arbitrarily. For example, \citet{gao_dynamic_2018} use \{0.005, 0.008, 0.012, 0.015, 0.02\} as input values for a threshold variable, and \{20, 40, 60, 80, 100\} as input for query range size. While these values may be representative for their given domain, this approach requires the authors to rationalize both the chosen ranges and the amount of values used. Furthermore, if a variable is varied in isolation, it is not possible to capture interactions between variables. For example, in \cite{ding_differentially_2011-4}, the authors vary the number of dimensions, while setting cardinality and \eptext\ to fixed values. As such the trend for error when varying the number of dimensions is just captured at a fixed setting.

Hence, we identify three existing problems: 1) the relationship between error and an algorithm's input may be unknown, 2) data oblivious error may result in incorrect error predictions, and 3) choosing representative values for empirical experiments is difficult. 
To mitigate these problems we propose a novel application of \textit{factor experiments}~\cite{noauthor_nistsematech_2013, nistsematech_511_whatis,nistsematech_5333_full-factorial-design}, a statistical approach, to the domain of differential privacy. Here, we show how empirical error measurements can be used to construct an error prediction model using (multiple) linear regression. As such, we are able to model the relationship between all input variables, including data, and error. Accordingly, for the example with \eptext\ and population as variables, the prediction model would be in the following format:
\begin{align}
    y=&\gamma_0+\gamma_{threshold} \times \textsf{threshold}+\gamma_{range} \times \textsf{range}\nonumber\\
    &+\gamma_{threshold:range}\times \textsf{threshold}:\textsf{range}
\end{align}
where $y$ is the predicted error for a specific setting, $\gamma_0$ is the intercept, \textsf{threshold} and \textsf{range} are \textit{coded value} representations of the factors, and \textsf{threshold:range} is the possible interaction between factors. Hence, the prediction model is able to predict the error for any value (within the model's span) of  \textsf{threshold} and \textsf{range}.

More importantly, factor experiments provide a systematic way to choose the experiment settings where the most information can be extracted. Consequently, our methodology tackles all of the three identified problems by 1) modeling the relationship between variables and error, 2) involving all input variables in model creation, and 3) minimizing the samples required, allowing for efficient experiments.

We expect our methodology to be valid for any differentially private algorithm: factor experiments allow both numerical and categorical variables, and the analyst may choose any suitable error metric for their domain. To put our methodology into context, we will conduct a case study. In our case study, we run a poll where the algorithm traverses a tree structure before delivering a differentially private reply. Now, we will argue that our use case is particularly interesting in the context of our methodology. First, we have noticed that it is difficult to model the error correctly due to allowing for arbitrarily complex tree structures, where we identify six variables that need to be varied in experiments. Next, it is also difficult to argue for what constitutes a 'good' experiment setting in this case. As such, we believe the many variables' effect on error in our particular use case is difficult to investigate using methods from the current literature. Accordingly, we use \Randori~\cite{nelson_randori_2021} as a use case where we create a prediction model for error. \Randori\ is a set of tools for gathering poll data under \textit{local differential privacy}~\cite{warner_randomized_1965}. So far, \Randori\ can predict error analytically through Chernoff bounds, but this error is not data aware. In this paper, we extend \Randori\ by adding a simulation tool where users can generate synthetic poll data and empirically evaluate error.

To summarize, prediction models created using our methodology will be able to answer the following questions:
\begin{itemize}
    \item What is each variable's impact/effect on error?
    \item Are there any relationships/interactions between variables?
\end{itemize}
Hence, our contribution is a method for constructing accuracy/error prediction models.

\section{Background}\label{paperE:sec:background}
In this paper, we join two well-known areas: differential privacy and \textit{statistical design of experiments} (DOE)~\cite{nistsematech_431_doe}. To provide the reader the necessary background, we describe the \tradeoff\ in differential privacy. As we expect our readers to mainly come from the area of differential privacy, we also introduce terminology used in DOE.

\subsection{Differential Privacy}\label{paperE:sec:differential-privacy}
Differential privacy~\cite{dwork_calibrating_2006,dwork_differential_2006} is a statistical notion of privacy that quantifies the privacy loss. Since differential privacy is a definition and not an implementation, differential privacy can be achieved in different ways, but must always satisfy \Cref{paperE:def:dp}. To define differential privacy, we must first define neighboring data sets (\Cref{paperE:def:neighbors}).

\begin{mydef}[Neighboring Data Sets]\label{paperE:def:neighbors}
Two data sets, $D$ and $D'$, are neighboring if and only if they differ on at most one element $d$. That is, $D'$ can be constructed from $D$ by adding or removing one single element $d$:
\begin{center}
    $D' = D \pm d$
\end{center}
\end{mydef}

\begin{mydef}[\eptext-Differential Privacy]\label{paperE:def:dp}
A randomized algorithm $f$ is \eptext-differentially private if for all neighboring data sets D, D' and for all sets of outputs $\mathcal{S}$\\
\[\text{Pr}[f(D) \in \mathcal{S}] \leq exp(\ep) \times \text{Pr}[f(D') \in \mathcal{S}]\]
where the probability is taken over the randomness of the algorithm $f$.
\end{mydef}
 
Although differential privacy gives strong mathematical privacy guarantees, implementations introduce some kind of error, relative to an exact but non-private algorithm, to achieve said privacy. The accuracy of a differentially private algorithm can be investigated through analytical accuracy bounds, such as Chernoff bounds. These analytical accuracy bounds are often expressed in general terms, i.e. they do not define error for a specific algorithm, such as the Chernoff bound given by \citet{kasiviswanathan_what_2011} in \Cref{paperE:def:accuracy}.

\begin{mydef}[($\alpha, \beta$)-usefulness]\label{paperE:def:accuracy}
Let $X$ be a random variable representing the error of the output of a differentially private algorithm $f'$, $n$ is the population size and $\alpha, \beta \in (0,\frac{1}{2})$, where $\beta = 2e^{-2\alpha^2n}$. Then with probability 1-$\beta$, the error $X$ is bounded by at most error $\alpha$: 
\begin{center}
$\text{Pr}[X \leq \alpha] \geq 1-\beta$
\end{center}
We say that $f'$ is ($\alpha, \beta$)-useful~\cite{zhu_differential_2017}.
\end{mydef}

Note that this formula in particular does not define how to express error. That is, error must be defined on a per-algorithm basis. For example, \citet{kasiviswanathan_what_2011} use misclassification error as their error metric. Still, the resulting accuracy bounds cover the entire \textit{possible} range of error the algorithm can achieve. That is, such theoretical accuracy bounds focus on the worst case error~\cite{hay_principled_2016}. In other words, the bounds do not describe how error is distributed within the bound. For example, high errors may have very low probability, but an analyst may still condemn the algorithm because the accuracy bounds are not tight enough. Consequently, predicting error using analytical methods can be overly pessimistic.

Furthermore, it can be difficult to properly model the error in order to construct a Chernoff bound. The data dependence of an algorithm's error is particularly important to capture. As \citet{hay_principled_2016} point out, a number of differentially private algorithms are indeed data dependent. Hence, data can have an impact on error, but the current literature offers no guidelines on modeling error correctly.

\subsection{Designed Experiments}\label{paperE:sec:factor-experiments}
In this paper, we will empirically measure the error of a differentially private algorithm. As a consequence, we need to plan and conduct experiments. More specifically, we will conduct \textit{factor experiments}~\cite{noauthor_nistsematech_2013, nistsematech_511_whatis}, which is a more efficient way of conducting experiments than changing \textit{one factor at a time} (OFAT)~\cite{fisher_statistical_1990}. Here, a factor is the same as a variable, and we will use these terms interchangeably.

\begin{figure}[htb]
    \centering
    \includegraphics[scale=0.4]{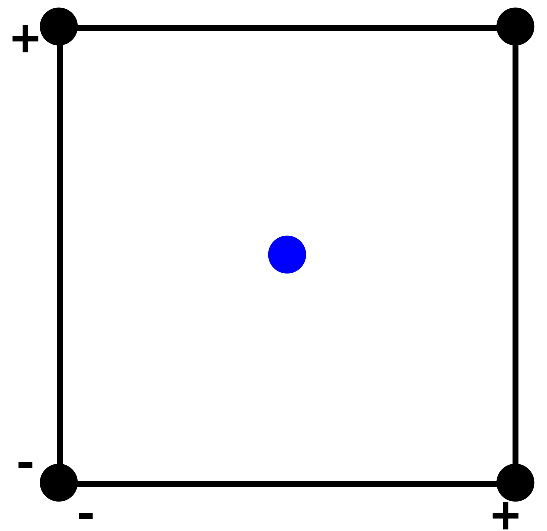}
    \caption{The space covered by a factor experiment with two factors. Black dots represents the factors at high/low respectively, and the blue dot is the baseline.}
    \label{paperE:fig:points-of-interest}
\end{figure}

With factor experiments, we are able to change several factors simultaneously, allowing us to run fewer experiments in total. Essentially, factor experiments is a way of designing experiments such that we can maximize what is learned given a fixed number of measurements~\cite{nistsematech_511_whatis}. For example, conducting an experiment with two different factors that each can take on 100 different values would require 10 000 measurements with the OFAT approach. Using these same factors but instead running \textit{two-level} factor experiments, we only need to measure the \textit{response} at each edge of the space. That is, only measurements from the black dots in \Cref{paperE:fig:points-of-interest} are required for factor experiments, whereas the response from each coordinate in the space is required using OFAT.

Hence, two-level factor experiments with two factors ($k=2$) require only $2^k=2^2= 4$ measurements. In summary, with two-level factor experiments, $2^k$ measurements are needed for an experiment with $k$ factors. Naturally, factor experiments are much more efficient than OFAT.

When running factor experiments, \textit{coded values} are used to denote \textit{actual values}. For example, two-level experiments usually have a low ('-' or '-1') and a high ('+' or '+1') coded value which are related to the actual values as follows:
\begin{align}
    v_{coded} &= \frac{v-a}{b},\\
\text{where }a &= \frac{v_{high} + v_{low}}{2},\\
b &= \frac{v_{high} - v_{low}}{2}
\end{align}
So, in a real use case, with the high value (+1) 1000, and the low value (-1) 100, the actual value 500 is represented by the coded value -$\frac{5}{45}$.

Another point of interest in factor experiments is the \textit{baseline}. The baseline is the center point (the blue dot in \Cref{paperE:fig:points-of-interest}) of the entire space that we cover. Consequently, the baseline always has the coded value 0 for each factor. 

Using the $2^k$ responses from the factor experiments, it is possible to construct a prediction model. In this paper, we will construct a linear prediction model using (multiple) linear regression. Given two factors $A$ and $B$, the linear model can be written as follows:
\begin{align}
    y=\gamma_0+\gamma_1 A+\gamma_2B+\gamma_{12}AB + \textsf{experimental error}
\end{align}
Where the constant $\gamma_0$ is the response at the baseline, and $AB$ is included to capture the possible \textit{interaction} between factor A and B.

Since the prediction model is linear, we will later show how to confirm these assumptions and validate the fit of the model. We also note that in case of non-linear systems, one can instead use three-level factorial designs~\cite{nistsematech_5339_nodate}, which are less efficient but are able to capture curvature.
\section{Methodology}\label{paperE:sec:method}
We propose a methodology consisting of four stages:
\begin{enumerate}
    \item Experiment design
    \item Data collection/generation
    \item Model creation
    \item Model validation
\end{enumerate}

After going through all the stages, the prediction model is ready to be used.

\subsection{Experiment Design}\label{paperE:sec:experiment-design}
We propose using two-level factor experiments. This allows linear prediction models to be created. Note that it is important to not choose maximum or minimum values for the levels, as such values likely will be too extreme and not produce a valid model~\cite{dunn_process_2021}. Instead, choose values that are feasible within the domain. Accordingly, the prediction model will be valid within the space the two levels span, but will not be able to make predictions for values outside. This step is necessary, as extreme values will likely break the assumptions about linearity that allow us to create a linear prediction model.

Next, the $k$ factors involved needs to be identified. This can be done in different ways. The authors note that in software systems, this process is much more straightforward than in for example physical systems, since all possible factors are represented in code. As such, it should be possible to extract all factors from the code directly.

In cases where there are many factors, it might be a good idea to run \textit{screening designs} first, using \textit{fractional designs}~\cite{nistsematech_5334_fractional-designs} experiments to reduce the number of measurements needed. Basically, a fractional design only includes some of the $2^k$ points, but are chosen in a systematic way. With screening designs, it is possible to determine if there are factors that can be ignored without running the full $2^k$ experiments.

\textbf{Our use case:}
In \Randori, data is gathered in poll format. A poll consists of a number of questions and a fixed set of answer alternatives. We represent these questions as trees where a node is either an answer alternative or a question. Furthermore, we also allow follow-up questions in our poll. As such, some answer alternatives have question nodes as children.

Answers to the poll are then gathered using \textit{randomized response}~\cite{warner_randomized_1965}. In randomized response, a respondent will answer truthfully with some probability, \textsf{Pr[truth]}, and will otherwise choose a random answer according to a known distribution. In \Randori, the known distribution is represented through weights attached to each answer alternative.

\begin{figure}[htb]
    \centering
    \includegraphics[width=\linewidth]{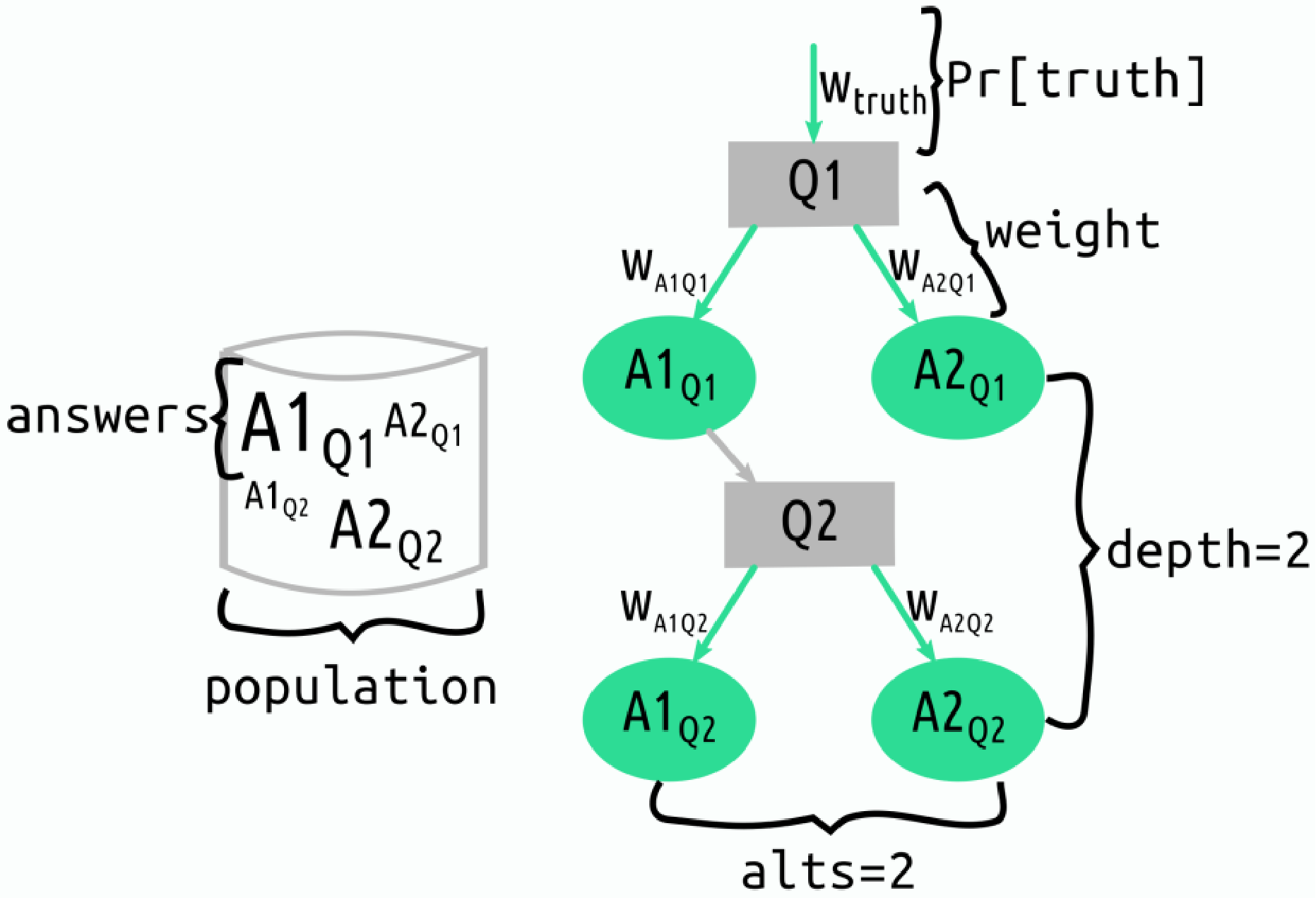}
    \caption{The factors used as input to \Randori, including both data (to the left) and algorithm parameters (to the right). Here, question nodes are gray and answer alternatives are green.}
    \label{paperE:fig:randori-factors}
\end{figure}

From our use case, we identify six factors to include in our experiment design. Here, \textsf{Pr[truth]} and \textsf{relative alternative weight} are due to randomized response. \textsf{Tree depth} and \textsf{Number of alternatives} are due to the poll's tree structure. Next, to make our model data aware, we include both the \textsf{Population} and the \textsf{Number of answers} which corresponds to the number of respondents that choose the answer alternative that we target in our measurements. We illustrate all of our identified factors in \Cref{paperE:fig:randori-factors}. When we measure the error, we will choose one of the alternatives as our target, for example $A1_{Q1}$.

In \Cref{paperE:tab:factors} we show all our factors and define the levels for each factor.

\begin{table}[htb]
\centering
\begin{tabular}{rlll}
Factor&+&- \\\toprule
\rowcolor{gray!10}\textsf{Pr[truth]}& High & Low \\
\textsf{Tree depth}& Deep & Shallow \\
\rowcolor{gray!10}\textsf{Number of alternatives} & Many & Few \\
\textsf{Relative alternative weight}& High & Low\\
\rowcolor{gray!10}\textsf{Population}& Many & Few \\
\textsf{Number of answers} & Many & Few \\\bottomrule
\end{tabular}
\caption{Factors, and their respective levels}
\label{paperE:tab:factors}
\end{table}

Now, it makes sense to explain why we have not included \eptext\ among our factors. In our case, one thing we want to investigate is the impact of the poll structure on the error. However, there is not a one-to-one mapping between \eptext\ and poll structure. That is, while \eptext\ can be calculated from the structure of the poll, different structures can result in the same value of \eptext. As such, only varying \eptext\ would not allow us to deduce a unique poll structure.

\subsection{Data Collection/Generation}
Data can either be collected from real experiments or generated synthetically. That is, responses from any differentially private algorithm can be used. Note that synthetic data does not make the model less valid: the prediction model will be valid for the entire space covered by the factors. In fact, if the algorithm can be simulated we recommend doing so, as this also eliminates the need to gather potentially sensitive data. Basically, the finesse of factor experiments is that we do not look to sample specific representative settings, but rather we want to be able to cover all values within a known space.

Since results from differentially private algorithms are probabilistic, it is also important to decide whether to measure an average error, or just one measurement per experiment setting. In this step, it is also important to decide which metric to use for error comparison. 

Next, create a table for all the possible combinations of the $k$ factors for a total of $2^k$ combinations. In physical systems, it is customary to produce the measurements in random order to avoid systematic errors.

\textbf{Our use case:}
We construct a tool where we can generate synthetic data and measure the empirical error introduced by randomized response. This tool simulates respondents answering a given poll on \Randori's format. We call this tool the \simenv.

We decide to run each setting 30 times, i.e. $n=30$, to measure the average error. We also decide to use \textit{mean average percentage error} (MAPE) as our error metric:
\begin{align}
   \text{MAPE} = \frac{1}{n}\sum_{t=1}^{n}\left\lvert\frac{x_t - x'_t}{x_t}\right\rvert \times 100 
\end{align}

Here, we will calculate the MAPE for one target answer alternative. As such, we measure the distance between the actual percentage ($x$) of respondents that chose the target alternative, and the estimated percentage ($x'$) calculated from the randomized responses. 

\subsection{Model Creation}
From the measured error, it is now possible to create the prediction model. The prediction model is calculated using (multiple) linear regression. To create the prediction model, we suggest using the programming language R. In R, pass the data to the \texttt{lm} function and R will output a model. This model will include the effect of each variable and all present interactions between variables.

\subsection{Model Validation}
To test the fit of the model, we first check that the assumptions about linearity hold. Next, the predictions made by the model also need to be investigated. That is, more measurements need to be gathered and compared to the model's predictions for the same setting.

If the model has a good fit, the \textit{residuals} will be small. We use the following formula to calculate the residual $r_i$ when comparing a prediction $y_i$ to a sample measurement $s_i$ for some coordinate $i$:
\begin{align}
    r_i = y_i - s_i
\end{align}
A numerical measurement of the model's fit is the (multiple) $R^2$, the coefficient of determination. A high value of $R^2$ is necessary but not sufficient for concluding that the fit is good~\cite{nistsematech_444_test-fit}. Next, compare the $R^2$ value to the adjusted $R^2$ (calculated as follows: $R_{adj}^2=1-(1-R^2)\frac{N-1}{N-p-1}$, where $N$ is the sample size and $p$ is the number of predictors). The value of $R^2$ and the adjusted $R^2$ should be close. Otherwise, a difference indicates that there are terms in the model that are not significant~\cite{nistsematech_544_testing-the-model}. Consequently, if $R^2$ and adjusted $R^2$ differ much, insignificant terms can be removed from the model. In this step, the programming language R can help with providing suggestions for which effects are significant.

Next, we recommend using visual methods to further validate the model due to NIST's recommendation~\cite{nistsematech_524_residual-behavior}. These visual methods allow conclusions to be drawn that cannot be drawn from merely observing $R^2$.

We suggest the following three visual methods:
\begin{enumerate}
    \item Histogram
    \item Residual vs. fitted plot
    \item Q-Q normal plot
\end{enumerate}

First, use a histogram to test the residuals for normality. Here, the residuals are expected to have the shape of a normal distribution, and to be centered around 0. 

Next, for the residual vs. fitted plot, values should be randomly scattered around 0 on the y-axis~\cite{nistsematech_524_residual-behavior}. We also expect the \textit{locally weighted scatterplot smoothing} (LOWESS)~\cite{nistsematech_lowess} curve to be flat, since this shows that a linear model is reasonable. 

Last, using the Q-Q normal plot shows if the residuals come from a common distribution as the prediction model. If the data sets come from common distributions, the points should be close to the plotted line.

\begin{figure}[htb]
    \centering
    \includegraphics[scale=0.4]{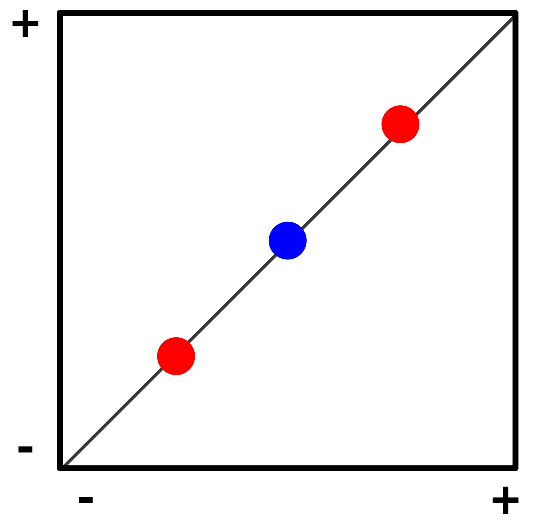}
    \caption{The center point, i.e. the baseline represented by the blue dot, and the red dots at (-0.5, -0.5) and (0.5,0.5) respectively}
    \label{paperE:fig:test-points}
\end{figure}

\textbf{Strategy if the model does not fit:} To get quick feedback about the model's fit, pick the three points in \Cref{paperE:fig:test-points}. Next, calculate the residuals for these points.

\begin{figure}[htb]
    \centering
    \includegraphics[scale=0.4]{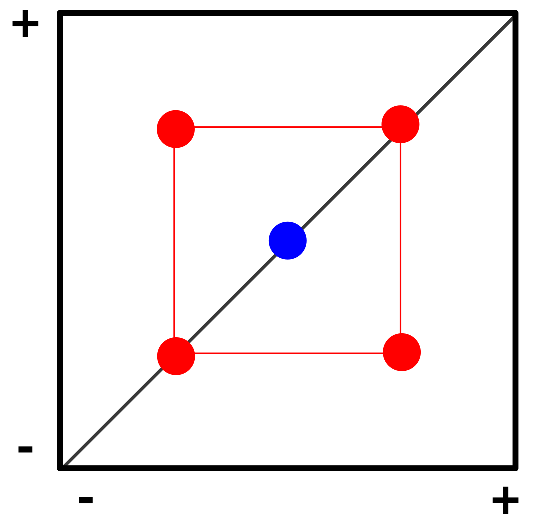}
    \caption{Adding the points (0.5,-0.5) and (-0.5,0.5) allows us to zoom in and find a new target space within the red lines}
    \label{paperE:fig:space}
\end{figure}

In cases where the residuals are high, re-use the samples from \Cref{paperE:fig:test-points} and add the remaining samples needed to create a new, smaller space. That is, systematically zoom in and target a smaller space to make the predictions on. We illustrate this new smaller space in 2D to be able to show a geometric explanation in \Cref{paperE:fig:space}.

\section{Results}\label{paperE:sec:results}
Next, we will apply our methodology to our use case where we estimate error for poll data. Here, we present the tool we used to generate data (the \simenv) and then we show how we iteratively apply the methodology to reach an adequate prediction model.

\subsection{Simulation Environment}\label{paperE:sec:simulation-environment}
\begin{figure}[htb]
\centering
\includegraphics[width=\linewidth]{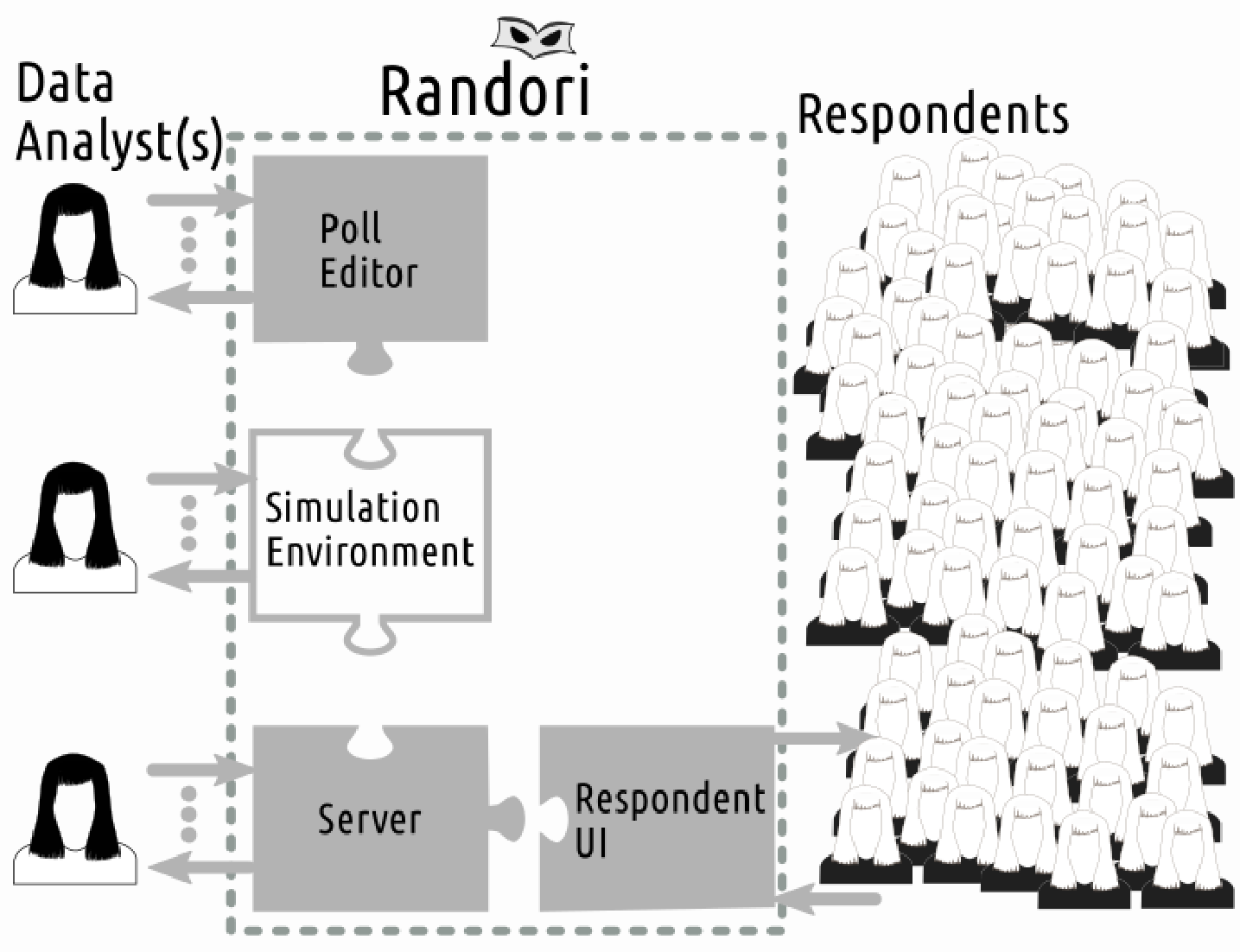}
\caption{The \simenv\ (white puzzle piece) in relation to existing \Randori\ tools}
\label{paperE:fig:randori-user-perspective}
\end{figure}

We have built a simulation environment using a Jupyter notebook~\cite{project_jupyter_project_2021} that takes input on a portable JSON format. The \simenv\ is an additional tool to the \Randori~\footnote{\url{https://github.com/niteo/randori}} (\Cref{paperE:fig:randori-user-perspective}) set of open source tools.

Here, a user can construct an arbitrarily complex poll using \Randori's \polleditor. Next, the poll can be imported into the \simenv\ where the user can tweak all the input variables. In addition, the \simenv\ is able to simulate the respondents' answers either based on probability distributions or a deterministic distribution, although we only use deterministic distributions in this paper.

\subsection{Experiments}\label{paperE:sec:experiment-setup}
We run a factor experiment with $k=6$, and calculate the error as MAPE. We run each experiment $n=30$ times.

Using the actual values in \Cref{paperE:tab:factors-values} we produce the measurements in \Cref{paperE:tab:result-experiment1} (in Appendix due to length). 

\begin{table}[htb]
\centering
\begin{tabular}{rlll}
Factor&Baseline&+1&-1 \\
\toprule
\rowcolor{gray!10}\textsf{Pr[truth]} & 50\% & 90\% & 10\% \\
\textsf{Tree depth} &3 & 5 & 1 \\
\rowcolor{gray!10}\textsf{Number of alternatives} & 6 & 10 & 2 \\
\textsf{Relative alternative weight} &50\% & 90\% & 10\%\\
\rowcolor{gray!10}\textsf{Population} & 50500 & 100 000 & 1000 \\
\textsf{Number of answers} &50\%& 90\% & 10\% \\\bottomrule
\end{tabular}
\caption{Factors and the actual values used for corresponding coded values. In the case of \weight\ and \pop\ the percentage is used for the target alternative, and the remainder is uniformly distributed among siblings.}
\label{paperE:tab:factors-values}
\end{table}

We enter our data in R and create a prediction model using the \texttt{lm} function. Calculating the residual for the baseline, we get a significant error of 384.6646. We pick two additional settings and measure them (\Cref{paperE:tab:residuals-exp1}) to convince ourselves that the model is indeed a bad fit. 

\begin{table}[htb]
\centering
\begin{tabular}{rlll}
Setting & $y_i$ & $s_i$ & $r_i$  \\\toprule
\rowcolor{gray!10}(0, 0, 0, 0, 0, 0) & 418.7087 & 34.04411  & 384.6646\\
(0.5, 0.5, 0.5, 0.5, 0.5, 0.5) & 124.8765 & 14.41732 & 110.4592 \\
\rowcolor{gray!10}(-0.5, -0.5, -0.5, -0.5, -0.5, -0.5) & 731.8813 & 38.23649 & 693.6448\\\bottomrule
\end{tabular}
\caption{Residuals calculated using the prediction model for the first experiment}
\label{paperE:tab:residuals-exp1}
\end{table}

As a result, we move on to sample the $2^6$ points that covers half the previous space i.e using the settings from \Cref{paperE:tab:factors-residual}. The measured MAPE is in \Cref{paperE:tab:result-experiment2} (in Appendix due to length). We then use these measurements to construct a new prediction model over the smaller space.

\begin{table}[htb]
\centering
\begin{tabular}{rlll}
Factor&Baseline&+0.5&-0.5 \\
\toprule
\rowcolor{gray!10}\textsf{Pr[truth]} & 50\% & 70\% & 30\% \\
\textsf{Tree depth} &3 & 4 & 2 \\
\rowcolor{gray!10}\textsf{Number of alternatives} & 6 & 8 & 4 \\
\textsf{Relative alternative weight} &50\% & 70\% & 30\%\\
\rowcolor{gray!10}\textsf{Population} & 50500 & 75750 & 25250 \\
\textsf{Number of answers} &50\%& 70\% & 30\% \\\bottomrule
\end{tabular}
\caption{Factors and the values used for calculating residuals}
\label{paperE:tab:factors-residual}
\end{table}

From entering our measured values into R's \texttt{lm} function, we get a model with 64 coefficients. Using the model, we notice that the prediction for the baseline has improved significantly. The updated prediction is $32.89371$, which gives us a residual of $34.04411-32.89371=1.1504$. Hence, we move on to validate our model.

\section{Analysis}\label{paperE:sec:analysis}
Next, we move on to validate our model according to our methodology. After validating the model, we will interpret the model.

\subsection{Evaluating the Model}
In order to validate the model, we need to investigate the behavior of the residuals. Hence, we need more measurements. We have decided to pick settings to sample from two sets:
\begin{enumerate}
    \item The corners ($2^6$ points) of the middle of the model (like in \Cref{paperE:fig:space}) and the center point
    \item Any coordinate in the space
\end{enumerate}

We randomly pick 20 points (except that we always include the center point in the first set) from each of the two approaches, giving us a total of 40 samples to calculate residuals from. Be aware that you may also need to adjust values in certain cases. In our case, we need to take into account that some of our factors are discrete. For example \depth\ is a discrete value and our corner values 0.25 and -0.25 would correspond to a depth of 3.5 and 2.5 respectively. Consequently, we chose to fix \depth\ to 3. The points and their corresponding MAPE is shown in \Cref{paperE:tab:validation-points}.

\begin{table}[htb]
\begin{tabularx}{\linewidth}{Xlllllll}
{} &  \footnotesize\truth &  \footnotesize\depth &  \footnotesize\alts &  \footnotesize\weight &  \footnotesize\textsf{pop} &  \footnotesize\answers &      MAPE \\
\toprule
\rowcolor{gray!10}0 & 0 & 0 & 0 & 0 & 0 & 0 & 34.04411\\
1  &   0.25 &   0.00 &  0.25 &    0.25 &        0.25 &     0.25 &  20.17603 \\
\rowcolor{gray!10}2  &  -0.25 &   0.00 &  0.25 &   -0.25 &        0.25 &    -0.25 &  48.18286 \\
3  &   0.25 &   0.00 & -0.25 &   -0.25 &        0.25 &    -0.25 &  31.06755 \\
\rowcolor{gray!10}4  &  -0.25 &   0.00 &  0.25 &   -0.25 &       -0.25 &     0.25 &  50.33476 \\
5  &   0.25 &   0.00 & -0.25 &    0.25 &        0.25 &     0.25 &  19.59611 \\
\rowcolor{gray!10}6  &  -0.25 &   0.00 &  0.25 &    0.25 &        0.25 &     0.25 &  27.66037 \\
7  &  -0.25 &   0.00 & -0.25 &   -0.25 &        0.25 &    -0.25 &  46.24753 \\
\rowcolor{gray!10}8  &  -0.25 &   0.00 & -0.25 &    0.25 &        0.25 &     0.25 &  26.60268 \\
9  &   0.25 &   0.00 & -0.25 &    0.25 &        0.25 &    -0.25 &  17.30670 \\
\rowcolor{gray!10}10 &  -0.25 &   0.00 &  0.25 &    0.25 &       -0.25 &    -0.25 &  25.07704 \\
11 &  -0.25 &   0.00 & -0.25 &   -0.25 &        0.25 &    -0.25 &  46.36067 \\
\rowcolor{gray!10}12 &  -0.25 &   0.00 & -0.25 &   -0.25 &        0.25 &    -0.25 &  46.18749 \\
13 &   0.25 &   0.00 & -0.25 &    0.25 &       -0.25 &     0.25 &  19.71108 \\
\rowcolor{gray!10}14 &   0.25 &   0.00 & -0.25 &   -0.25 &       -0.25 &     0.25 &  33.26383 \\
15 &  -0.25 &   0.00 &  0.25 &   -0.25 &       -0.25 &    -0.25 &  48.09976 \\
\rowcolor{gray!10}16 &  -0.25 &   0.00 &  0.25 &    0.25 &       -0.25 &     0.25 &  27.58968 \\
17 &  -0.25 &   0.00 & -0.25 &    0.25 &        0.25 &    -0.25 &  22.55290 \\
\rowcolor{gray!10}18 &  -0.25 &   0.00 &  0.25 &    0.25 &        0.25 &    -0.25 &  24.97823 \\
19 &   0.25 &   0.00 & -0.25 &    0.25 &        0.25 &     0.25 &  19.61443 \\\midrule
\rowcolor{gray!10}20 &  -0.50 &  -0.50 & -0.50 &    0.03 &       -0.46 &     0.28 &   8.42964 \\ 
21 &   0.16 &   0.25 &  0.00 &    0.32 &       -0.25 &     0.38 &  28.34642 \\
\rowcolor{gray!10}22 &  -0.06 &  -0.25 & -0.50 &    0.03 &       -0.31 &    -0.32 &   8.82148 \\
23 &  -0.50 &   0.25 & -0.25 &    0.03 &        0.03 &    -0.29 &  53.20864 \\
\rowcolor{gray!10}24 &   0.21 &   0.50 &  0.00 &    0.12 &       -0.17 &     0.34 &  36.71494 \\
25 &   0.31 &   0.50 &  0.25 &    0.34 &       -0.02 &     0.39 &  29.04886 \\
\rowcolor{gray!10}26 &  -0.49 &   0.25 &  0.25 &   -0.22 &       -0.12 &     0.07 &  63.40224 \\
27 &  -0.27 &  -0.50 &  0.00 &    0.35 &        0.29 &     0.34 &  65.43967 \\
\rowcolor{gray!10}28 &   0.39 &   0.25 &  0.50 &    0.21 &       -0.03 &     0.38 &  25.73380 \\
29 &   0.39 &  -0.25 &  0.00 &    0.30 &        0.13 &     0.28 &   3.46581 \\
\rowcolor{gray!10}30 &  -0.45 &   0.50 &  0.50 &    0.06 &       -0.04 &    -0.21 &  59.91642 \\
31 &  -0.00 &   0.50 & -0.25 &   -0.36 &        0.05 &    -0.02 &  47.62934 \\
\rowcolor{gray!10}32 &  -0.20 &  -0.25 & -0.50 &   -0.03 &        0.16 &     0.42 &  21.80034 \\
33 &  -0.14 &   0.25 &  0.50 &   -0.40 &        0.11 &     0.46 &  53.57877 \\
\rowcolor{gray!10}34 &   0.11 &   0.00 & -0.25 &   -0.48 &       -0.35 &    -0.21 &  39.38831 \\
35 &   0.14 &   0.00 &  0.00 &   -0.37 &        0.15 &     0.02 &  38.41253 \\
\rowcolor{gray!10}36 &  -0.09 &  -0.50 & -0.50 &   -0.41 &       -0.47 &    -0.39 &   5.75857 \\
37 &  -0.19 &   0.50 &  0.25 &   -0.08 &        0.44 &    -0.19 &  52.70103 \\
\rowcolor{gray!10}38 &   0.42 &  -0.50 & -0.25 &   -0.19 &        0.00 &    -0.01 &   2.18997 \\
39 &  -0.47 &   0.50 & -0.25 &    0.33 &       -0.33 &     0.35 &  51.42151 \\
\bottomrule
\end{tabularx}
\caption{The sampled points used and their measured MAPE}
    \label{paperE:tab:validation-points}
\end{table}

%"R² is the proportion of the variance in the dependent variable that is predictable from the independent variable(s)"
First, we check the value of our $R^2$. For our model, the $R^2$ is 0.8419. However, we notice that the adjusted $R^2$ is significantly lower, 0.5929. Seeing as we have 64 coefficients, it seems reasonable to simplify our model to avoid overfitting. We update our model in R to only involve the effects that R marks as significant. To do this, we enter the suggested effects in R, which in our case are: \texttt{lm(formula = MAPE $\sim$ truth + alts + weight + truth*depth+depth*weight + truth*depth*weight + depth*weight*answers )}. Now, we end up with a $R^2$ of 0.7846, and an adjusted $R^2$ of 0.7562. These values are still high, and since they are now significantly closer, we move on to validate the model visually.

\begin{figure}[htb]
    \centering
    \includegraphics[width=\linewidth]{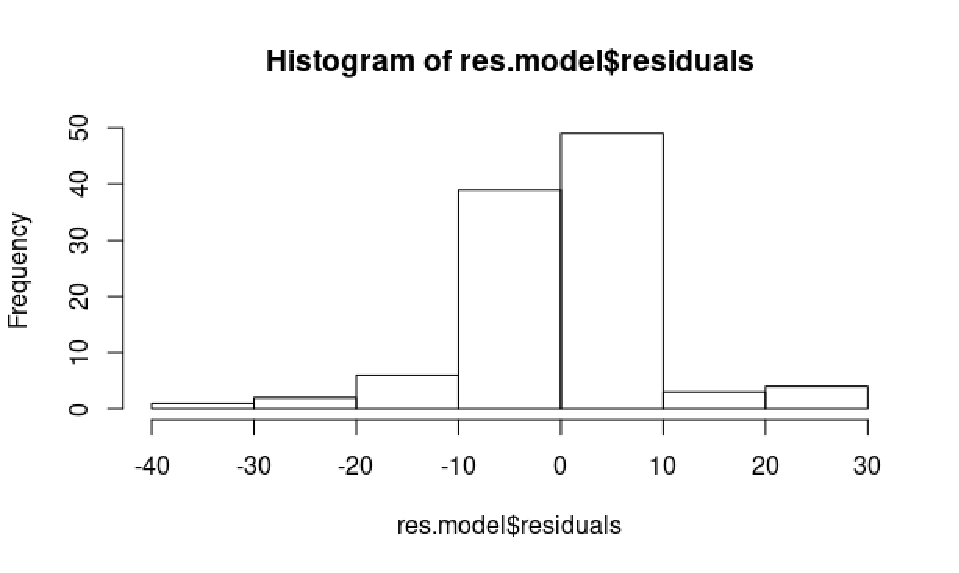}
    \caption{A histogram of the residuals}
    \label{paperE:fig:residual-hist}
\end{figure}

Next, we plot the residuals as a histogram in~\Cref{paperE:fig:residual-hist}. From the histogram, we see that our residuals are indeed centered around 0. The histogram indicates a normal distribution. Hence, we move on to the next test.

\begin{figure}[htb]
    \centering
    \includegraphics[width=\linewidth]{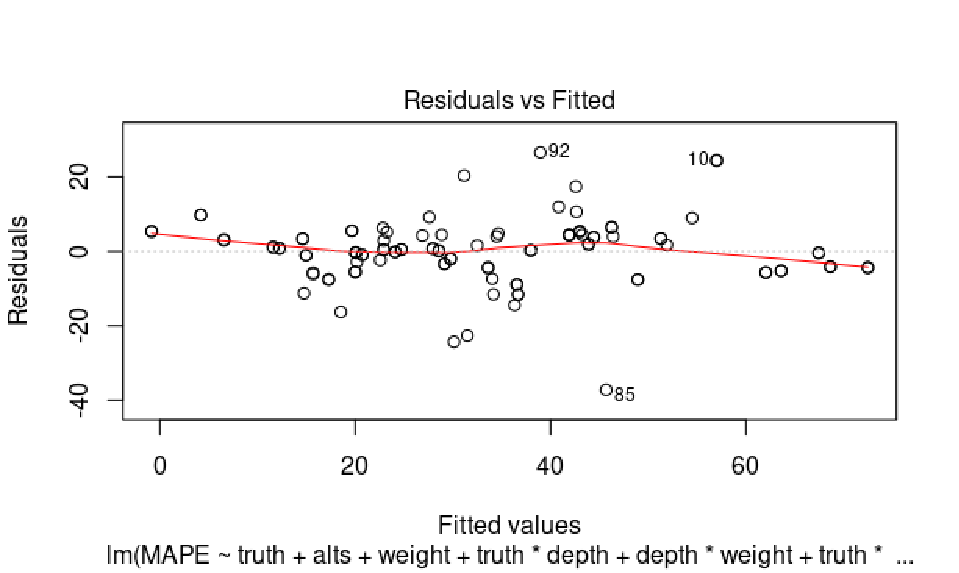}
    \caption{Residuals represented as circles, fitted values as the dotted line. The red line represents the LOWESS fit of the residuals vs. fitted values.}
    \label{paperE:fig:fitted-vs-residual}
\end{figure}

Now, we want to investigate the relationship between fitted values (measurements) and the model's prediction. Then, we plot fitted values vs. predictions in \Cref{paperE:fig:fitted-vs-residual}. We observe that the residuals appear to not have a specific shape around the y-axis. We also conclude that the LOWESS fit curve appears to be almost flat.

\begin{figure}[htb]
    \centering
    \includegraphics[width=\linewidth]{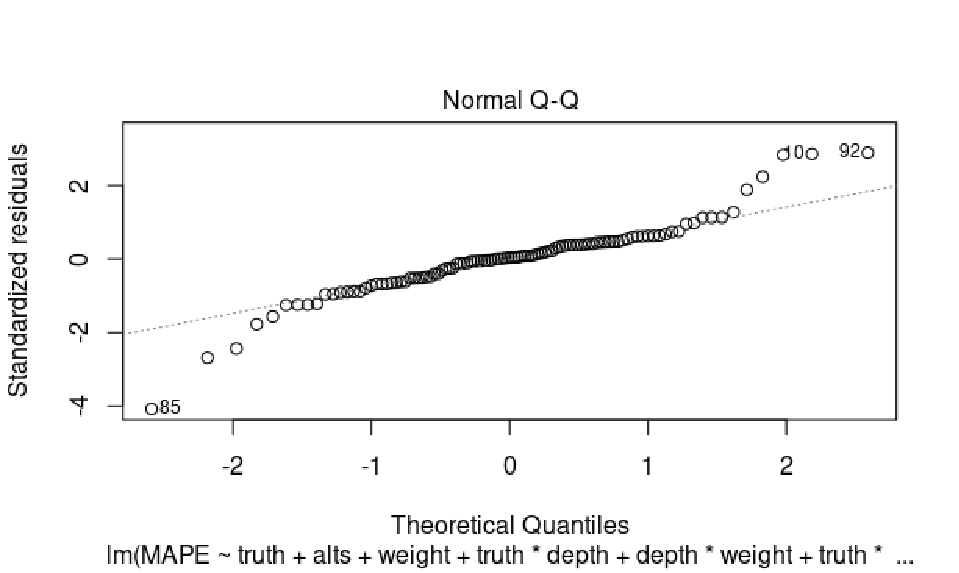}
    \caption{The normal quantile-quantile plot}
    \label{paperE:fig:normal-q-q}
\end{figure}

Finally, we investigate the normal Q-Q plot (\Cref{paperE:fig:normal-q-q}). We see that most points follow the plotted line, indicating that our predictions come from the same distribution as the measured values. Hence, we conclude that our prediction model is valid for our use case.

\subsection{Interpreting the Model}
The model is now ready to be used. That is, any value within each factor's range [high,low] can be plugged in to produce an error prediction. It is also possible to set $y \leq c$, with $c$ being our maximum tolerable error, and then find which settings satisfy the inequality. Our final error prediction model is as follows:
\begin{align}
    y &= 32.501266 - 29.023493\times\truth\ + 5.037411\times\alts\ \nonumber  \\
    &- 16.562410\times\weight\ + 1.449934\times\depth\ \nonumber\\
    &+ 1.856916\times\answers\ + 10.044302\times\truth:\depth\ \nonumber\\
    &- 28.397984\times\weight:\depth\ \nonumber\\
    &+ 4.175231\times\truth:\weight\ \nonumber\\
    &+ 8.535667\times\depth:\answers\ \nonumber\\ 
    &- 8.402531\times\weight:\answers\ \nonumber\\
    &+ 51.134829\times\truth:\weight:\depth\ \nonumber\\
    &+ 25.945740\times\weight:\depth:\answers 
\end{align}
We note that the simplification step has allowed us to completely eliminate \pop\ from our factors. As such, we draw the conclusion that the population size itself does not have a significant impact on error.

\begin{figure}[htb]
    \centering
    \includegraphics[width=\linewidth]{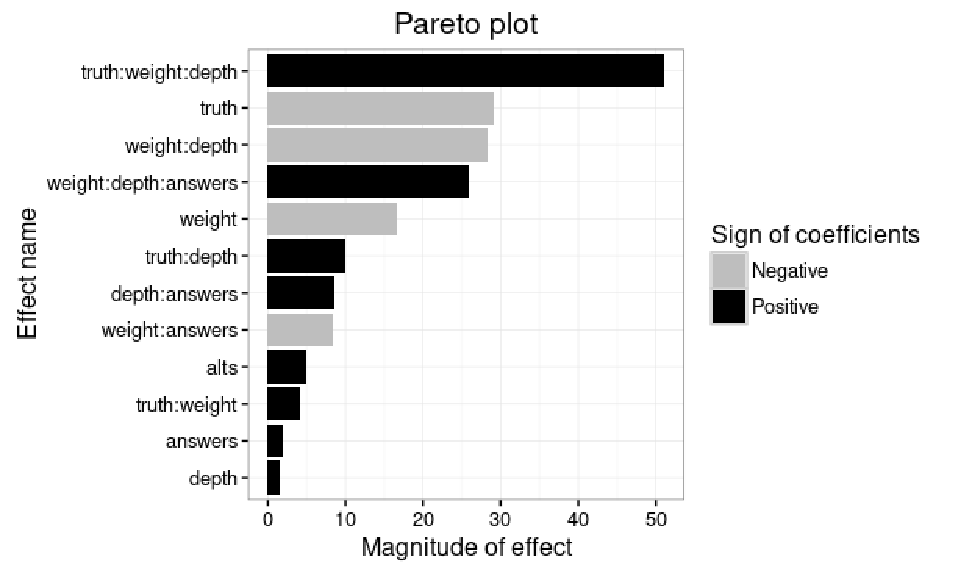}
    \caption{The Pareto plot of the simplified model}
    \label{paperE:fig:pareto-simplified}
\end{figure}

To get an overview of our model, we use a Pareto plot~\cite{rdocumentation_paretochart_nodate} (\Cref{paperE:fig:pareto-simplified}) which allows us to visually compare all effects at once. Here, effects are ordered by magnitude.

From the plot, it is clear that \truth:\weight:\depth\ affects error the most. Maybe most notably, \truth:\weight:\depth\ increases error whereas its components \truth\ and \weight:\depth\ both decrease error. From examining the Pareto plot, it seems that \truth:\weight\ is the interaction that causes the increase in error.

As expected, \truth\ has a negative impact on error. That is, a high value of \truth\ will reduce error. More surprisingly, \truth\ is involved in several interactions which all increase error.

It may be tempting to completely ignore \answers\ and \depth\ as these two factors have the lowest magnitude of effect. However, ignoring these factors is dangerous: they are both involved in interactions that have significantly higher magnitude.

The factor \alts\ is the only one that does not have interactions. It may seem counter-intuitive that having more siblings have such a small impact on error. Still, the magnitude of this effect may very well be due to our choice to input polls where we uniformly distribute the remaining weight among the siblings.

Hence, we can give \Randori's users the following advice: use the model to find local minima or ranges. The model can also be used to find minima and ranges while accounting for known constraints such as for example $\textsf{Pr[truth]} \leq 0.5$. When working in \Randori's \polleditor\ it is important to beware of the main effect \truth:\weight:\depth\ and its building blocks. As \weight\ primarily is involved in decreasing error, we recommend increasing \weight\ before tweaking the other factors.

\section{Discussion, Limitations and Future Work}\label{paperE:sec:discussion}
A limitation in our work is that the prediction models we create are linear. As such, prediction can be off in cases where the error is in fact non-linear. Still, factor experiments can nevertheless be used to make predictions for non-linear systems. To facilitate for non-linear systems the factor levels have to be chosen differently: i.e. we would need 3 levels~\cite{nistsematech_5339_nodate} instead of 2. Hence, our approach can be adapted to create non-linear models by running more experiments.

Additionally, we know that error should also depend on the, non-linear, term $exp(\ep)$ from the definition of differential privacy. Still, it is not clear how the term $exp(\ep)$ and other, algorithm specific, factors compare in order of magnitude. As such, more research is needed to see if \eptext\ can be modeled in a suitable way, or if perhaps \eptext\ needs to be transformed to be linear ($ln(exp(\ep)$). Nevertheless, factor experiments still provide a systematic and efficient way to explore the impact of different variables on error. That is, factor experiments may still be used to explore the other factors' impact on error. Hence, while it may not always be possible to extract an accurate prediction model, factor experiments are still useful when determining which data points should be used as input to test the accuracy of a differentially private algorithm.

Furthermore, factor experiments provide a possible way to systematically predict error for \textit{all} representative input data sets for a differentially private algorithm. That is, instead of using real data sets to predict error, factor experiments statistically emulate all possible data sets bounded by the experiment's levels (the high/low values for each variable in our case). Hence, using factor experiments to create prediction models can be more robust statistically than making predictions based on one real data set.

Whether the model is correct or not will be identified when testing the model according to our methodology. If the model is incorrect it can be due to error being non-linear, but it can also be due to not including all relevant factors. As such, an incorrect model requires further investigation.

Accordingly, correctly identifying relevant factors is crucial to building a correct model. Still, there exists no recognized way of correctly and efficiently identifying all factors. As mentioned in \Cref{paperE:sec:experiment-design}, it is nonetheless possible to try if a factor is relevant using \textit{screening designs} before running a full factorial experiment. From our use case, it is nonetheless clear that some candidate factors rule themselves out by simply being impossible to implement. For example, we considered having the factor \textit{number of parent siblings} together with \depth, which results in the impossible combination of having no parents (\depth=0) and also having parent siblings. Hence, we believe looking for possible contradictions among factors is important when designing the experiments.

In order to not create contradicting factors, we have also decided to only model the \weight\ for the target alternative. That is, we set the weight for the target alternative (or the target's parent), and uniformly divide the remainder among the siblings. For example, when a target has weight 70\% and three siblings, each sibling gets $\frac{100-70}{3}$\% each. As such, we have not investigated settings where the siblings have non-uniform weight distributions.

One decision that may seem controversial is that we do not include \eptext\ as one of the factors in our model. While we do not tweak \eptext\ directly, we do in fact adjust \eptext\ by changing the structure of the poll. The reason we have chosen to indirectly tweak \eptext\ as to tweaking it directly is that one single value of \eptext\ corresponds to multiple poll structures, whereas one poll structure corresponds to exactly one value of \eptext. Hence, while it may seem unintuitive at first, indirectly tweaking \eptext\ makes more sense than tweaking it directly in our case.

Somewhat surprising is that population was eliminated from our prediction model in the simplification step. We argue that the elimination of population is because \answers\ is related to \pop\ (the probability of choosing some alternative $Ai_{Qj}$ is \textsf{Pr[$Ai_{Qj}$]} = \pop*\answers), and population therefore becomes redundant. It is also possible that the choice of error measurement, MAPE in our case, contributes to making population irrelevant since it is a relative measurement of error as opposed to an absolute measurement.

Finally, we note that in this paper we have measured the error of leaf nodes in a tree. Still, with the known relationships between answers, it appears to be possible to further post-process and add accuracy to parent answers. We believe including the number of children as a factor would be an interesting path to explore next in order to better understand the effect of this post-processing. Put differently, the challenge here is properly modeling the factors without creating contradictions between factors.
\section{Related Work}\label{paperE:sec:related-work}
As mentioned in \Cref{paperE:sec:introduction}, evaluating error empirically is not a new topic within differential privacy. However, creating prediction models from empirical data appears to be a novel approach.

The work closest to ours is \dpbench~\cite{hay_principled_2016}, which is an error evaluation framework for differentially private algorithms. In \dpbench, the authors propose a set of \textit{evaluation principles}, including guidelines for creating diverse input for algorithms. Hence, \dpbench\ has a strong focus on understanding the data-dependence of an algorithm's error. Still, \dpbench\ does not produce an error prediction model like we do, nor does it minimize the number of experiments needed to conduct.

We also note that \dpcomp~\cite{hay_exploring_2016} is the closest work to our \simenv. \dpcomp\ allows users to compare how the accuracy of a differentially private algorithm is affected by varying input data. Our work is similar in the sense that our \simenv\ also is intended to be used to evaluate \tradeoff s. Our \simenv\ is also inspired by \dpbench's evaluation principles and consequently allows data following different distributions to be entered and evaluated. However, our simulation environment is less general than \dpcomp, since our solution uses one fixed algorithm.
\section{Conclusion}\label{paperE:sec:conclusion}
We have presented a methodology for empirically estimating error in differentially private algorithms which 1) models the relationships between input parameters, 2) is data aware, and 3) minimizes the measurements required as input. Hence, prediction models created using our methodology allow for expressive, data aware, error prediction. Moreover, we conducted a case study where we apply our methodology to a setting where error is measured from poll structures. To support our use case, we have added a simulation tool to the \Randori\ open source tool suite, adding the functionality of generating synthetic data and evaluating error empirically.

From our case study, we were able to create a prediction model for error using six factors. After evaluating and simplifying our model, we are able to answer the two questions from our introduction. First, there are 13 main effects on error. Next, there are seven interactions.

From evaluating the prediction model we found that our model has a good fit. As such, our novel application of factor experiments shows promising results as a methodology for error evaluation of differentially private algorithms.

Consequently, we have contributed with a novel application of a methodology that shows promise for error prediction of differentially private algorithms. In addition, we have also built a simulation environment that generates synthetic poll data and measures error through simulating randomized response.

One interesting path for future work is to investigate if, and how, the number of factors used in the model prediction affects the model's fit. Along a similar line of thought, it would also be interesting to attempt to create prediction models for well known differentially private algorithms and libraries. As such, we encourage the use of our methodology in order to construct error prediction models for other differentially private algorithms.

\begin{acks}
This work was partly funded by the Swedish Foundation for Strategic Research (SSF) and the Swedish Research Council (VR).
\end{acks}
%\balance
\bibliographystyle{ACM-Reference-Format}
\bibliography{ref.bib}


\begin{thebibliography}{30}

%%% the Web address.

\ifx \showCODEN    \undefined \def \showCODEN     #1{\unskip}     \fi
\ifx \showDOI      \undefined \def \showDOI       #1{#1}\fi
\ifx \showISBNx    \undefined \def \showISBNx     #1{\unskip}     \fi
\ifx \showISBNxiii \undefined \def \showISBNxiii  #1{\unskip}     \fi
\ifx \showISSN     \undefined \def \showISSN      #1{\unskip}     \fi
\ifx \showLCCN     \undefined \def \showLCCN      #1{\unskip}     \fi
\ifx \shownote     \undefined \def \shownote      #1{#1}          \fi
\ifx \showarticletitle \undefined \def \showarticletitle #1{#1}   \fi
\ifx \showURL      \undefined \def \showURL       {\relax}        \fi
% invisible to TeX
\providecommand\bibfield[2]{#2}
\providecommand\bibinfo[2]{#2}
\providecommand\natexlab[1]{#1}
\providecommand\showeprint[2][]{arXiv:#2}

\bibitem[\protect\citeauthoryear{Benkhelif, Fessant, Cl{\'e}rot, and
  Raschia}{Benkhelif et~al\mbox{.}}{2017}]%
        {benkhelif_co-clustering_2017-3}
\bibfield{author}{\bibinfo{person}{Tarek Benkhelif}, \bibinfo{person}{Fran{\c
  c}oise Fessant}, \bibinfo{person}{Fabrice Cl{\'e}rot}, {and}
  \bibinfo{person}{Guillaume Raschia}.} \bibinfo{year}{2017}\natexlab{}.
\newblock \showarticletitle{Co-Clustering for Differentially Private Synthetic
  Data Generation}. In \bibinfo{booktitle}{\emph{Personal Analytics and
  Privacy. {{An}} Individual and Collective Perspective}},
  \bibfield{editor}{\bibinfo{person}{Riccardo Guidotti}, \bibinfo{person}{Anna
  Monreale}, \bibinfo{person}{Dino Pedreschi}, {and} \bibinfo{person}{Serge
  Abiteboul}} (Eds.). \bibinfo{publisher}{{Springer International Publishing}},
  \bibinfo{address}{{Cham}}, \bibinfo{pages}{36--47}.
\newblock
\showISBNx{978-3-319-71970-2}


\bibitem[\protect\citeauthoryear{Chen, Shen, and Jin}{Chen
  et~al\mbox{.}}{2015}]%
        {chen_private_2015}
\bibfield{author}{\bibinfo{person}{Rui Chen}, \bibinfo{person}{Yilin Shen},
  {and} \bibinfo{person}{Hongxia Jin}.} \bibinfo{year}{2015}\natexlab{}.
\newblock \showarticletitle{Private {{Analysis}} of {{Infinite Data Streams}}
  via {{Retroactive Grouping}}}. In \bibinfo{booktitle}{\emph{Proceedings of
  the 24th ACM International on Conference on Information and Knowledge
  Management}} \emph{(\bibinfo{series}{CIKM '15})}.
  \bibinfo{publisher}{Association for Computing Machinery},
  \bibinfo{address}{New York, NY, USA}, \bibinfo{pages}{1061–1070}.
\newblock
\showISBNx{9781450337946}
\urldef\tempurl%
\url{https://doi.org/10.1145/2806416.2806454}
\showDOI{\tempurl}


\bibitem[\protect\citeauthoryear{Ding, Winslett, Han, and Li}{Ding
  et~al\mbox{.}}{2011}]%
        {ding_differentially_2011-4}
\bibfield{author}{\bibinfo{person}{Bolin Ding}, \bibinfo{person}{Marianne
  Winslett}, \bibinfo{person}{Jiawei Han}, {and} \bibinfo{person}{Zhenhui Li}.}
  \bibinfo{year}{2011}\natexlab{}.
\newblock \showarticletitle{Differentially {{Private Data Cubes}}: {{Optimizing
  Noise Sources}} and {{Consistency}}}. In
  \bibinfo{booktitle}{\emph{Proceedings of the 2011 {{ACM SIGMOD International
  Conference}} on {{Management}} of {{Data}}}}
  \emph{(\bibinfo{series}{{{SIGMOD}} '11})}. \bibinfo{publisher}{{ACM}},
  \bibinfo{address}{{New York, NY, USA}}, \bibinfo{pages}{217--228}.
\newblock
\showISBNx{978-1-4503-0661-4}
\urldef\tempurl%
\url{https://doi.org/10.1145/1989323.1989347}
\showDOI{\tempurl}


\bibitem[\protect\citeauthoryear{Dunn}{Dunn}{2021}]%
        {dunn_process_2021}
\bibfield{author}{\bibinfo{person}{Kevin Dunn}.}
  \bibinfo{year}{2021}\natexlab{}.
\newblock \bibinfo{title}{Process {{Improvement Using Data}}}.
\newblock
\newblock
\newblock
\shownote{Release 0f428b.}


\bibitem[\protect\citeauthoryear{Dwork}{Dwork}{2006}]%
        {dwork_differential_2006}
\bibfield{author}{\bibinfo{person}{Cynthia Dwork}.}
  \bibinfo{year}{2006}\natexlab{}.
\newblock \showarticletitle{Differential {{Privacy}}}.
\newblock In \bibinfo{booktitle}{\emph{Automata, {{Languages}} and
  {{Programming}}}}, \bibfield{editor}{\bibinfo{person}{Michele Bugliesi},
  \bibinfo{person}{Bart Preneel}, \bibinfo{person}{Vladimiro Sassone}, {and}
  \bibinfo{person}{Ingo Wegener}} (Eds.). Number 4052 in
  \bibinfo{series}{Lecture {{Notes}} in {{Computer Science}}}.
  \bibinfo{publisher}{{Springer}}, \bibinfo{address}{{Berlin, Heidelberg}},
  \bibinfo{pages}{1--12}.
\newblock
\showISBNx{978-3-540-35907-4 978-3-540-35908-1}


\bibitem[\protect\citeauthoryear{Dwork, McSherry, Nissim, and Smith}{Dwork
  et~al\mbox{.}}{2006}]%
        {dwork_calibrating_2006}
\bibfield{author}{\bibinfo{person}{Cynthia Dwork}, \bibinfo{person}{Frank
  McSherry}, \bibinfo{person}{Kobbi Nissim}, {and} \bibinfo{person}{Adam
  Smith}.} \bibinfo{year}{2006}\natexlab{}.
\newblock \showarticletitle{Calibrating Noise to Sensitivity in Private Data
  Analysis}. In \bibinfo{booktitle}{\emph{Theory of Cryptography}},
  \bibfield{editor}{\bibinfo{person}{Shai Halevi} {and} \bibinfo{person}{Tal
  Rabin}} (Eds.). \bibinfo{publisher}{{Springer}}, \bibinfo{address}{{Berlin,
  Heidelberg}}, \bibinfo{pages}{265--284}.
\newblock
\showISBNx{978-3-540-32732-5}


\bibitem[\protect\citeauthoryear{Fisher}{Fisher}{1990}]%
        {fisher_statistical_1990}
\bibfield{author}{\bibinfo{person}{Ronald~A. Fisher}.}
  \bibinfo{year}{1990}\natexlab{}.
\newblock \bibinfo{booktitle}{\emph{Statistical Methods, Experimental Design,
  and Scientific Inference} (\bibinfo{edition}{1st edition} ed.)}.
\newblock \bibinfo{publisher}{{Oxford University Press}},
  \bibinfo{address}{{Oxford, England}}.
\newblock
\showISBNx{978-0-19-852229-4}


\bibitem[\protect\citeauthoryear{{Gao} and {Ma}}{{Gao} and {Ma}}{2018}]%
        {gao_dynamic_2018}
\bibfield{author}{\bibinfo{person}{Ruichao {Gao}} {and} \bibinfo{person}{Xuebin
  {Ma}}.} \bibinfo{year}{2018}\natexlab{}.
\newblock \showarticletitle{Dynamic Data Histogram Publishing Based on
  Differential Privacy}. In \bibinfo{booktitle}{\emph{2018 IEEE Intl Conf on
  Parallel Distributed Processing with Applications, Ubiquitous Computing
  Communications, Big Data Cloud Computing, Social Computing Networking,
  Sustainable Computing Communications
  (ISPA/IUCC/BDCloud/SocialCom/SustainCom)}}. \bibinfo{publisher}{IEEE},
  \bibinfo{address}{Melbourne, VIC, Australia}, \bibinfo{pages}{737--743}.
\newblock
\urldef\tempurl%
\url{https://doi.org/10.1109/BDCloud.2018.00111}
\showDOI{\tempurl}


\bibitem[\protect\citeauthoryear{Hay, Machanavajjhala, Miklau, Chen, and
  Zhang}{Hay et~al\mbox{.}}{2016a}]%
        {hay_principled_2016}
\bibfield{author}{\bibinfo{person}{Michael Hay}, \bibinfo{person}{Ashwin
  Machanavajjhala}, \bibinfo{person}{Gerome Miklau}, \bibinfo{person}{Yan
  Chen}, {and} \bibinfo{person}{Dan Zhang}.} \bibinfo{year}{2016}\natexlab{a}.
\newblock \showarticletitle{Principled Evaluation of Differentially Private
  Algorithms Using DPBench}. In \bibinfo{booktitle}{\emph{Proceedings of the
  2016 International Conference on Management of Data}}
  \emph{(\bibinfo{series}{SIGMOD '16})}. \bibinfo{publisher}{Association for
  Computing Machinery}, \bibinfo{address}{New York, NY, USA},
  \bibinfo{pages}{139–154}.
\newblock
\showISBNx{9781450335317}
\urldef\tempurl%
\url{https://doi.org/10.1145/2882903.2882931}
\showDOI{\tempurl}


\bibitem[\protect\citeauthoryear{Hay, Machanavajjhala, Miklau, Chen, Zhang, and
  Bissias}{Hay et~al\mbox{.}}{2016b}]%
        {hay_exploring_2016}
\bibfield{author}{\bibinfo{person}{Michael Hay}, \bibinfo{person}{Ashwin
  Machanavajjhala}, \bibinfo{person}{Gerome Miklau}, \bibinfo{person}{Yan
  Chen}, \bibinfo{person}{Dan Zhang}, {and} \bibinfo{person}{George Bissias}.}
  \bibinfo{year}{2016}\natexlab{b}.
\newblock \showarticletitle{Exploring {{Privacy}}-{{Accuracy Tradeoffs Using
  DPComp}}}. In \bibinfo{booktitle}{\emph{Proceedings of the 2016
  {{International Conference}} on {{Management}} of {{Data}}}}
  \emph{(\bibinfo{series}{{{SIGMOD}} '16})}. \bibinfo{publisher}{{ACM}},
  \bibinfo{address}{{New York, NY, USA}}, \bibinfo{pages}{2101--2104}.
\newblock
\showISBNx{978-1-4503-3531-7}
\urldef\tempurl%
\url{https://doi.org/10.1145/2882903.2899387}
\showDOI{\tempurl}


\bibitem[\protect\citeauthoryear{Kasiviswanathan, Lee, Nissim, Raskhodnikova,
  and Smith}{Kasiviswanathan et~al\mbox{.}}{2011}]%
        {kasiviswanathan_what_2011}
\bibfield{author}{\bibinfo{person}{Shiva Kasiviswanathan},
  \bibinfo{person}{Homin Lee}, \bibinfo{person}{Kobbi Nissim},
  \bibinfo{person}{Sofya Raskhodnikova}, {and} \bibinfo{person}{Adam Smith}.}
  \bibinfo{year}{2011}\natexlab{}.
\newblock \showarticletitle{What {{Can We Learn Privately}}?}
\newblock \bibinfo{journal}{\emph{SIAM J. Comput.}} \bibinfo{volume}{40},
  \bibinfo{number}{3} (\bibinfo{date}{Jan.} \bibinfo{year}{2011}),
  \bibinfo{pages}{793--826}.
\newblock
\showISSN{0097-5397}
\urldef\tempurl%
\url{https://doi.org/10.1137/090756090}
\showDOI{\tempurl}


\bibitem[\protect\citeauthoryear{Li, Cui, Meng, and Ma}{Li
  et~al\mbox{.}}{2019}]%
        {li_ihp_2019}
\bibfield{author}{\bibinfo{person}{Hui Li}, \bibinfo{person}{Jiangtao Cui},
  \bibinfo{person}{Xue Meng}, {and} \bibinfo{person}{Jianfeng Ma}.}
  \bibinfo{year}{2019}\natexlab{}.
\newblock \showarticletitle{{{IHP}}: {{Improving}} the Utility in Differential
  Private Histogram Publication}.
\newblock \bibinfo{journal}{\emph{Distributed and Parallel Databases}}
  \bibinfo{volume}{37} (\bibinfo{year}{2019}), \bibinfo{pages}{721--750}.
\newblock


\bibitem[\protect\citeauthoryear{Li, Xiong, Jiang, and Liu}{Li
  et~al\mbox{.}}{2015}]%
        {li_differentially_2015-9}
\bibfield{author}{\bibinfo{person}{Haoran Li}, \bibinfo{person}{Li Xiong},
  \bibinfo{person}{Xiaoqian Jiang}, {and} \bibinfo{person}{Jinfei Liu}.}
  \bibinfo{year}{2015}\natexlab{}.
\newblock \showarticletitle{Differentially Private Histogram Publication for
  Dynamic Datasets: An Adaptive Sampling Approach}. In
  \bibinfo{booktitle}{\emph{Proceedings of the 24th ACM International on
  Conference on Information and Knowledge Management}}
  \emph{(\bibinfo{series}{CIKM '15})}. \bibinfo{publisher}{Association for
  Computing Machinery}, \bibinfo{address}{New York, NY, USA},
  \bibinfo{pages}{1001–1010}.
\newblock
\showISBNx{9781450337946}
\urldef\tempurl%
\url{https://doi.org/10.1145/2806416.2806441}
\showDOI{\tempurl}


\bibitem[\protect\citeauthoryear{Nelson}{Nelson}{2021}]%
        {nelson_randori_2021}
\bibfield{author}{\bibinfo{person}{Boel Nelson}.}
  \bibinfo{year}{2021}\natexlab{}.
\newblock \bibinfo{title}{Randori: {{Local Differential Privacy}} for {{All}}}.
\newblock
\newblock
\newblock
\shownote{arXiv:2101.11502 [cs].}


\bibitem[\protect\citeauthoryear{{NIST/SEMATECH}}{{NIST/SEMATECH}}{2013}]%
        {nistsematech_lowess}
\bibfield{author}{\bibinfo{person}{{NIST/SEMATECH}}.}
  \bibinfo{year}{2013}\natexlab{}.
\newblock \bibinfo{title}{4.1.4.4. {{LOESS}} (Aka {{LOWESS}})}.
\newblock
  \bibinfo{howpublished}{https://www.itl.nist.gov/div898/handbook/pmd/section1/pmd144.htm}.
\newblock
\newblock
\shownote{[Accessed: 2021-05-15].}


\bibitem[\protect\citeauthoryear{NIST/SEMATECH}{NIST/SEMATECH}{2013}]%
        {nistsematech_431_doe}
\bibfield{author}{\bibinfo{person}{NIST/SEMATECH}.}
  \bibinfo{year}{2013}\natexlab{}.
\newblock \bibinfo{title}{4.3.1. {{What}} Is Design of Experiments ({{DOE}})?}
\newblock
  \bibinfo{howpublished}{https://www.itl.nist.gov/div898/handbook/pmd/section3/\\pmd31.htm}.
\newblock
\newblock
\shownote{[Accessed: 2021-03-02].}


\bibitem[\protect\citeauthoryear{{NIST/SEMATECH}}{{NIST/SEMATECH}}{2013a}]%
        {nistsematech_444_test-fit}
\bibfield{author}{\bibinfo{person}{{NIST/SEMATECH}}.}
  \bibinfo{year}{2013}\natexlab{a}.
\newblock \bibinfo{title}{4.4.4. {{How}} Can {{I}} Tell If a Model Fits My
  Data?}
\newblock
  \bibinfo{howpublished}{https://www.itl.nist.gov/div898/handbook/pmd/section4/\\pmd44.htm}.
\newblock
\newblock
\shownote{[Accessed: 2021-02-24].}


\bibitem[\protect\citeauthoryear{{NIST/SEMATECH}}{{NIST/SEMATECH}}{2013b}]%
        {nistsematech_511_whatis}
\bibfield{author}{\bibinfo{person}{{NIST/SEMATECH}}.}
  \bibinfo{year}{2013}\natexlab{b}.
\newblock \bibinfo{title}{5.1.1. {{What}} Is Experimental Design?}
\newblock
  \bibinfo{howpublished}{https://www.itl.nist.gov/div898/handbook//pri/section1/\\pri11.htm}.
\newblock
\newblock
\shownote{[Accessed: 2021-02-17].}


\bibitem[\protect\citeauthoryear{NIST/SEMATECH}{NIST/SEMATECH}{2013}]%
        {nistsematech_524_residual-behavior}
\bibfield{author}{\bibinfo{person}{NIST/SEMATECH}.}
  \bibinfo{year}{2013}\natexlab{}.
\newblock \bibinfo{title}{5.2.4. {{Are}} the Model Residuals Well-Behaved?}
\newblock
  \bibinfo{howpublished}{https://www.itl.nist.gov/div898/handbook/pri/section2/\\pri24.htm}.
\newblock
\newblock
\shownote{[Accessed: 2021-02-17].}


\bibitem[\protect\citeauthoryear{{NIST/SEMATECH}}{{NIST/SEMATECH}}{2013a}]%
        {nistsematech_5333_full-factorial-design}
\bibfield{author}{\bibinfo{person}{{NIST/SEMATECH}}.}
  \bibinfo{year}{2013}\natexlab{a}.
\newblock \bibinfo{title}{5.3.3.3. {{Full}} Factorial Designs}.
\newblock
  \bibinfo{howpublished}{https://www.itl.nist.gov/div898/handbook/pri/section3\\/pri333.htm}.
\newblock
\newblock
\shownote{[Accessed: 2021-02-24].}


\bibitem[\protect\citeauthoryear{{NIST/SEMATECH}}{{NIST/SEMATECH}}{2013b}]%
        {nistsematech_5334_fractional-designs}
\bibfield{author}{\bibinfo{person}{{NIST/SEMATECH}}.}
  \bibinfo{year}{2013}\natexlab{b}.
\newblock \bibinfo{title}{5.3.3.4. {{Fractional}} Factorial Designs}.
\newblock
  \bibinfo{howpublished}{https://www.itl.nist.gov/div898/handbook/pri/section3\\/pri334.htm}.
\newblock
\newblock
\shownote{[Accessed: 2021-02-24].}


\bibitem[\protect\citeauthoryear{NIST/SEMATECH}{NIST/SEMATECH}{2013a}]%
        {nistsematech_5339_nodate}
\bibfield{author}{\bibinfo{person}{NIST/SEMATECH}.}
  \bibinfo{year}{2013}\natexlab{a}.
\newblock \bibinfo{title}{5.3.3.9. {{Three}}-Level Full Factorial Designs}.
\newblock
  \bibinfo{howpublished}{https://www.itl.nist.gov/div898/handbook//pri/section3/pri339.htm}.
\newblock
\newblock
\shownote{[Accessed: 2021-05-15].}


\bibitem[\protect\citeauthoryear{NIST/SEMATECH}{NIST/SEMATECH}{2013b}]%
        {nistsematech_544_testing-the-model}
\bibfield{author}{\bibinfo{person}{NIST/SEMATECH}.}
  \bibinfo{year}{2013}\natexlab{b}.
\newblock \bibinfo{title}{5.4.4. {{How}} to Test and Revise {{DOE}} Models}.
\newblock
  \bibinfo{howpublished}{https://www.itl.nist.gov/div898/handbook/pri/section4/\\pri44.htm}.
\newblock
\newblock
\shownote{[Accessed: 2021-02-24].}


\bibitem[\protect\citeauthoryear{{NIST/SEMATECH}}{{NIST/SEMATECH}}{2013}]%
        {noauthor_nistsematech_2013}
\bibfield{author}{\bibinfo{person}{{NIST/SEMATECH}}.}
  \bibinfo{year}{2013}\natexlab{}.
\newblock \bibinfo{title}{{{NIST}}/{{SEMATECH}} e-{{Handbook}} of {{Statistical
  Methods}}}.
\newblock
  \bibinfo{howpublished}{https://www.itl.nist.gov/div898/handbook/index.htm}.
\newblock
\newblock
\shownote{[Accessed: 2021-02-17].}


\bibitem[\protect\citeauthoryear{{Project Jupyter}}{{Project Jupyter}}{2021}]%
        {project_jupyter_project_2021}
\bibfield{author}{\bibinfo{person}{{Project Jupyter}}.}
  \bibinfo{year}{2021}\natexlab{}.
\newblock \bibinfo{title}{Project {{Jupyter}}}.
\newblock \bibinfo{howpublished}{https://www.jupyter.org}.
\newblock
\newblock
\shownote{[Accessed: 2021-05-15].}


\bibitem[\protect\citeauthoryear{{RDocumentation}}{{RDocumentation}}{2021}]%
        {rdocumentation_paretochart_nodate}
\bibfield{author}{\bibinfo{person}{{RDocumentation}}.}
  \bibinfo{year}{2021}\natexlab{}.
\newblock \bibinfo{title}{Pareto.Chart Function - {{RDocumentation}}}.
\newblock
  \bibinfo{howpublished}{https://www.rdocumentation.org/packages/qcc/versions/2.6/topics/pareto.chart}.
\newblock
\newblock
\shownote{[Accessed: 2021-05-15].}


\bibitem[\protect\citeauthoryear{Vadhan}{Vadhan}{2017}]%
        {vadhan_complexity_2017}
\bibfield{author}{\bibinfo{person}{Salil Vadhan}.}
  \bibinfo{year}{2017}\natexlab{}.
\newblock \showarticletitle{The {{Complexity}} of {{Differential Privacy}}}.
\newblock In \bibinfo{booktitle}{\emph{Tutorials on the {{Foundations}} of
  {{Cryptography}}: {{Dedicated}} to {{Oded Goldreich}}}},
  \bibfield{editor}{\bibinfo{person}{Yehuda Lindell}} (Ed.).
  \bibinfo{publisher}{{Springer International Publishing}},
  \bibinfo{address}{{Cham}}, \bibinfo{pages}{347--450}.
\newblock
\showISBNx{978-3-319-57048-8}
\urldef\tempurl%
\url{https://doi.org/10.1007/978-3-319-57048-8_7}
\showDOI{\tempurl}


\bibitem[\protect\citeauthoryear{Warner}{Warner}{1965}]%
        {warner_randomized_1965}
\bibfield{author}{\bibinfo{person}{Stanley~L. Warner}.}
  \bibinfo{year}{1965}\natexlab{}.
\newblock \showarticletitle{Randomized {{Response}}: {{A Survey Technique}} for
  {{Eliminating Evasive Answer Bias}}}.
\newblock \bibinfo{journal}{\emph{J. Amer. Statist. Assoc.}}
  \bibinfo{volume}{60}, \bibinfo{number}{309} (\bibinfo{date}{March}
  \bibinfo{year}{1965}), \bibinfo{pages}{63--69}.
\newblock
\showISSN{0162-1459}
\urldef\tempurl%
\url{https://doi.org/10.1080/01621459.1965.10480775}
\showDOI{\tempurl}


\bibitem[\protect\citeauthoryear{Xiao, Gardner, and Xiong}{Xiao
  et~al\mbox{.}}{2012}]%
        {xiao_dpcube_2012-1}
\bibfield{author}{\bibinfo{person}{Yonghui Xiao}, \bibinfo{person}{James
  Gardner}, {and} \bibinfo{person}{Li Xiong}.} \bibinfo{year}{2012}\natexlab{}.
\newblock \showarticletitle{{{DPCube}}: {{Releasing Differentially Private Data
  Cubes}} for {{Health Information}}}. In
  \bibinfo{booktitle}{\emph{International {{Conference}} on {{Data
  Engineering}} ({{ICDE}})}}. \bibinfo{publisher}{{IEEE}},
  \bibinfo{address}{Arlington, VA, USA}, \bibinfo{pages}{1305--1308}.
\newblock
\showISSN{1063-6382}
\urldef\tempurl%
\url{https://doi.org/10.1109/ICDE.2012.135}
\showDOI{\tempurl}


\bibitem[\protect\citeauthoryear{Zhu, Li, Zhou, and Yu}{Zhu
  et~al\mbox{.}}{2017}]%
        {zhu_differential_2017}
\bibfield{author}{\bibinfo{person}{Tianqing Zhu}, \bibinfo{person}{Gang Li},
  \bibinfo{person}{Wanlei Zhou}, {and} \bibinfo{person}{Philip~S. Yu}.}
  \bibinfo{year}{2017}\natexlab{}.
\newblock \bibinfo{booktitle}{\emph{Differential {{Privacy}} and
  {{Applications}}}}. \bibinfo{series}{Advances in {{Information Security}}},
  Vol.~\bibinfo{volume}{69}.
\newblock \bibinfo{publisher}{{Springer International Publishing}},
  \bibinfo{address}{{Cham}}.
\newblock
\showISBNx{978-3-319-62002-2 978-3-319-62004-6}
\urldef\tempurl%
\url{https://doi.org/10.1007/978-3-319-62004-6}
\showDOI{\tempurl}


\end{thebibliography}
\appendix
\section{Experiments}
\onecolumn

\footnotesize{
\begin{longtable}{cccccccc} 
Standard order&Pr[truth]&Tree depth&Number of alternatives&Alternative weight&Population&Number of answers& MAPE\\\toprule \endhead
N/A & 0& 0& 0& 0& 0& 0& 34.04411\\\midrule
\rowcolor{gray!10}1& -& -& -& -& -& -& 87.08667\\
2& +& -& -& -& -& -& 3.49111\\
\rowcolor{gray!10}3& -& +& -& -& -& -& 37.57905\\
4& +& +& -& -& -& -& 4.90007\\
\rowcolor{gray!10}5& -& -& +& -& -& -& 47.75\\
6& +& -& +& -& -& -& 6.58\\
\rowcolor{gray!10}7& -& +& +& -& -& -& 76.73124\\
8& +& +& +& -& -& -& 8.56657\\
\rowcolor{gray!10}9& -& -& -& +& -& -& 7365.33667\\
10&+& -& -& +& -& -& 96.20333\\
\rowcolor{gray!10}11&-& +& -& +& -& -& 1228.76234\\
12&+& +& -& +& -& -& 19.77456\\
\rowcolor{gray!10}13&-& -& +& +& -& -& 1456.40333\\
14&+& -& +& +& -& -& 18.47\\
\rowcolor{gray!10}15&-& +& +& +& -& -& 405.1528\\
16&+& +& +& +& -& -& 3.74374\\
\rowcolor{gray!10}17&-& -& -& -& +& -& 90.03673\\
18&+& -& -& -& +& -& 1.21997\\
\rowcolor{gray!10}19&-& +& -& -& +& -& 39.38121\\
20&+& +& -& -& +& -& 4.38645\\
\rowcolor{gray!10}21&-& -& +& -& +& -& 47.13567\\
22&+& -& +& -& +& -& 7.02496\\
\rowcolor{gray!10}23&-& +& +& -& +& -& 75.60747\\
24&+& +& +& -& +& -& 8.34256\\
\rowcolor{gray!10}25&-& -& -& +& +& -& 7362.4095\\
26&+& -& -& +& +& -& 98.25777\\
\rowcolor{gray!10}27&-& +& -& +& +& -& 1240.11986\\
28&+& +& -& +& +& -& 19.7394\\
\rowcolor{gray!10}29&-& -& +& +& +& -& 1466.18583\\
30&+& -& +& +& +& -& 18.8858\\
\rowcolor{gray!10}31&-& +& +& +& +& -& 403.33846\\
32&+& +& +& +& +& -& 4.16551\\
\rowcolor{gray!10}33&-& -& -& -& -& +& 61.83111\\
34&+& -& -& -& -& +& 8.08626\\
\rowcolor{gray!10}35&-& +& -& -& -& +& 88.29154\\
36&+& +& -& -& -& +& 9.66657\\
\rowcolor{gray!10}37&-& -& +& -& -& +& 63.75222\\
38&+& -& +& -& -& +& 8.2323\\
\rowcolor{gray!10}39&-& +& +& -& -& +& 89.69907\\
40&+& +& +& -& -& +& 10.02583\\
\rowcolor{gray!10}41&-& -& -& +& -& +& 811.41556\\
42&+& -& -& +& -& +& 10.13037\\
\rowcolor{gray!10}43&-& +& -& +& -& +& 310.01569\\
44&+& +& -& +& -& +& 2.16437\\
\rowcolor{gray!10}45&-& -& +& +& -& +& 738.71667\\
46&+& -& +& +& -& +& 9.07111\\
\rowcolor{gray!10}47&-& +& +& +& -& +& 300.02957\\
48&+& +& +& +& -& +& 2.1328\\
\rowcolor{gray!10}49&-& -& -& -& +& +& 61.99979\\
50&+& -& -& -& +& +& 7.9004\\
\rowcolor{gray!10}51&-& +& -& -& +& +& 88.42618\\
52&+& +& -& -& +& +& 9.84616\\
\rowcolor{gray!10}53&-& -& +& -& +& +& 63.75659\\
54&+& -& +& -& +& +& 7.95395\\
\rowcolor{gray!10}55&-& +& +& -& +& +& 89.82931\\
56&+& +& +& -& +& +& 9.95786\\
\rowcolor{gray!10}57&-& -& -& +& +& +& 810.22851\\
58&+& -& -& +& +& +& 9.99809\\
\rowcolor{gray!10}59&-& +& -& +& +& +& 310.55943\\
60&+& +& -& +& +& +& 2.44021\\
\rowcolor{gray!10}61&-& -& +& +& +& +& 737.21517\\
62&-& +& +& +& +& +& 299.99379\\
\rowcolor{gray!10}63&+& -& +& +& +& +& 9.01693\\
64&+& +& +& +& +& +& 2.20558\\\bottomrule
\caption{MAPE measurements for the experiment using -1 and +1 as coded value inputs}
\label{paperE:tab:result-experiment1}
\end{longtable}

\begin{longtable}{cccccccc} 
Standard order & Pr[truth] & Tree depth & Number of alternatives & Alternative weight & Population & Number of answers & MAPE\\\toprule \endhead
N/A & 0 & 0 & 0 & 0 & 0 & 0 & 34.04411\\\midrule
\rowcolor{gray!10}1 & - & - & - & - & - & - & 38.23649\\
2 & + & - & - & - & - & - & 17.89185\\
\rowcolor{gray!10}3 & - & + & - & - & - & - & 58.33831\\
4 & + & + & - & - & - & - & 25.18673\\
\rowcolor{gray!10}5 & - & - & + & - & - & - & 48.15875\\
6 & + & - & + & - & - & - & 25.15229\\
\rowcolor{gray!10}7 & - & + & + & - & - & - & 64.44095\\
8 & + & + & + & - & - & - & 27.66351\\
\rowcolor{gray!10}9 & - & - & - & + & - & - & 81.467\\
10 & + & - & - & + & - & - & 13.00362\\
\rowcolor{gray!10}11 & - & + & - & + & - & - & 9.89232\\
12 & + & + & - & + & - & - & 9.41709\\
\rowcolor{gray!10}13 & - & - & + & + & - & - & 56.28555\\
14 & + & - & + & + & - & - & 9.56171\\
\rowcolor{gray!10}15 & - & + & + & + & - & - & 19.75423\\
16 & + & + & + & + & - & - & 12.79737\\
\rowcolor{gray!10}17 & - & - & - & - & + & - & 38.11988\\
18 & + & - & - & - & + & - & 17.97198\\
\rowcolor{gray!10}19 & - & + & - & - & + & - & 58.37657\\
20 & + & + & - & - & + & - & 25.14935\\
\rowcolor{gray!10}21 & - & - & + & - & + & - & 48.43102\\
22 & + & - & + & - & + & - & 25.08915\\
\rowcolor{gray!10}23 & - & + & + & - & + & - & 64.49147\\
24 & + & + & + & - & + & - & 27.73975\\
\rowcolor{gray!10}25 & - & - & - & + & + & - & 81.24882\\
26 & + & - & - & + & + & - & 13.02403\\
\rowcolor{gray!10}27 & - & + & - & + & + & - & 9.5234\\
28 & + & + & - & + & + & - & 9.65797\\
\rowcolor{gray!10}29 & - & - & + & + & + & - & 56.3261\\
30 & + & - & + & + & + & - & 9.79661\\
\rowcolor{gray!10}31 & - & + & + & + & + & - & 19.70136\\
32 & + & + & + & + & + & - & 12.57202\\
\rowcolor{gray!10}33 & - & - & - & - & - & + & 52.6255\\
34 & + & - & - & - & - & + & 23.3408\\
\rowcolor{gray!10}35 & - & + & - & - & - & + & 66.96285\\
36 & + & + & - & - & - & + & 28.56059\\
\rowcolor{gray!10}37 & - & - & + & - & - & + & 54.63909\\
38 & + & - & + & - & - & + & 28.61188\\
\rowcolor{gray!10}39 & - & + & + & - & - & + & 68.09695\\
40 & + & + & + & - & - & + & 29.17961\\
\rowcolor{gray!10}41 & - & - & - & + & - & + & 45.78992\\
42 & + & - & - & + & - & + & 4.45637\\
\rowcolor{gray!10}43 & - & + & - & + & - & + & 23.87327\\
44 & + & + & - & + & - & + & 13.78785\\
\rowcolor{gray!10}45 & - & - & + & + & - & + & 41.37552\\
46 & + & - & + & + & - & + & 13.85628\\
\rowcolor{gray!10}47 & - & + & + & + & - & + & 25.68611\\
48 & + & + & + & + & - & + & 14.47902\\
\rowcolor{gray!10}49 & - & - & - & - & + & + & 52.71001\\
50 & + & - & - & - & + & + & 23.2522\\
\rowcolor{gray!10}51 & - & + & - & - & + & + & 66.94767\\
52 & + & + & - & - & + & + & 28.70839\\
\rowcolor{gray!10}53 & - & - & + & - & + & + & 54.66564\\
54 & + & - & + & - & + & + & 28.71268\\
\rowcolor{gray!10}55 & - & + & + & - & + & + & 68.04705\\
56 & + & + & + & - & + & + & 29.16309\\
\rowcolor{gray!10}57 & - & - & - & + & + & + & 45.72794\\
58 & + & - & - & + & + & + & 4.47782\\
\rowcolor{gray!10}59 & - & + & - & + & + & + & 23.84796\\
60 & + & + & - & + & + & + & 13.90072\\
\rowcolor{gray!10}61 & - & - & + & + & + & + & 41.23229\\
62 & - & + & + & + & + & + & 25.70945\\
\rowcolor{gray!10}63 & + & - & + & + & + & + & 13.88817\\
64 & + & + & + & + & + & + & 14.41732\\\bottomrule
\caption{MAPE measurements for the experiment using -0.5 and +0.5 as coded value inputs}
\label{paperE:tab:result-experiment2}
\end{longtable}
}

\end{document}